# Disk fragmentation in high-mass star formation

## High-resolution observations towards AFGL 2591-VLA 3

S. Suri[1], H. Beuther[1], C. Gieser[1], A. Ahmadi[2], Á. Sánchez-Monge[3], J. M. Winters[4], H. Linz[1], Th. Henning[1], M. T. Beltrán[5], F. Bosco[1], R. Cesaroni[5], T. Csengeri[6], S. Feng[7,8,9], M. G. Hoare[10] K. G. Johnston,[10] P. Klaassen[11], R. Kuiper[12,13], S. Leurini[14], S. Longmore[15], S. Lumsden[10], L. Maud[16], L. Moscadelli[5], T. Möller[3], A. Palau[17], T. Peters[18], R. E. Pudritz[19], S. E. Ragan[20], D. Semenov[1,21], P. Schilke[3], J. S. Urquhart[22], F. Wyrowski[23], H. Zinnecker[24]

*(Affiliations can be found after the references)*

**ABSTRACT**

*Context.* Increasing evidence suggests that, similar to their low-mass counterparts, high-mass stars form through a disk-mediated accretion process. At the same time, formation of high-mass stars still necessitates high accretion rates, and hence, high gas densities, which in turn can cause disks to become unstable against gravitational fragmentation.

*Aims.* We study the kinematics and fragmentation of the disk around the high-mass star forming region AFGL 2591-VLA 3 which was hypothesized to be fragmenting based on the observations that show multiple outflow directions.

*Methods.* We use a new set of high-resolution (0.″19) IRAM/NOEMA observations at 843 μm towards VLA 3 which allow us to resolve its disk, characterize the fragmentation, and study its kinematics. In addition to the 843 μm continuum emission, our spectral setup targets warm dense gas and outflow tracers such as HCN, $HC_3N$ and $SO_2$, as well as vibrationally excited HCN lines.

*Results.* The high resolution continuum and line emission maps reveal multiple fragments with subsolar masses within the inner ∼1000 AU of VLA 3. Furthermore, the velocity field of the inner disk observed at 843 μm shows a similar behavior to that of the larger scale velocity field studied in the CORE project at 1.37 mm.

*Conclusions.* We present the first observational evidence for disk fragmentation towards AFGL 2591-VLA 3, a source that was thought to be a single high-mass core. While the fragments themselves are low-mass, the rotation of the disk is dominated by the protostar with a mass of $10.3\pm1.8\ M_\odot$. These data also show that NOEMA Band 4 can obtain the highest currently achievable spatial resolution at (sub-)mm wavelengths in observations of strong northern sources.

## 1. Introduction

High-mass stars ($M > 8\ M_\odot$) regulate the dynamical and chemical evolution of the interstellar medium (ISM), yet the exact routes that lead to their formation are still under debate. The reason why observations of high-mass star formation run into a bottleneck is twofold. Firstly, these objects are rarer than their low-mass counterparts (Kroupa 2002; Chabrier 2003) and mostly found further away in dense clusters (e.g. Zinnecker & Yorke 2007; Beuther et al. 2007; Tan et al. 2014). Secondly, the lifetime of high-mass stars, due to their higher luminosities, is shorter because they burn their fuel much more rapidly. Because they live shorter lives, it is harder to observe them at a given stage. Despite these difficulties, there has been immense observational and theoretical effort focused towards understanding the stages of their formation and evolution.

At the molecular cloud scales, recent work reveals that filament fragmentation (e.g. Kainulainen et al. 2017) and mass accretion towards high- and intermediate-mass star-forming clumps (e.g. Hacar et al. 2018) play a key role in providing a terrain where such objects can form. When it comes to the collapse of individual cores in sub-parsec scales, theoretical work predicts formation of accretion disks around the central objects. A critical aspect of the high-mass star formation is that the Kelvin-Helmholtz time scale[1] of the protostars is much shorter than the free-fall time scale of their envelopes (e.g. Palla & Stahler 1993; Schilke 2015). This implies that high-mass protostars lack a pre-main sequence phase and instead, they ignite hydrogen while they are still in the accretion phase. Due to radiation pressure that kicks in early-on during their formation, the way high-mass stars can acquire masses above $40M_\odot$ has long been a puzzle (e.g. Wolfire & Cassinelli 1987). The most recent studies show, however, that with the existence of accretion disks, similar to those of low-mass stars, and asymmetric accretion, the accretion of material onto the high-mass cores is not hindered due to radiation pressure (e.g. Krumholz et al. 2009; Kuiper et al. 2010; Tan et al. 2014; Klassen et al. 2016; Kuiper & Hosokawa 2018).

Observations of disks around high-mass stars are scarce when compared to their low- and intermediate-mass counterparts, limiting our ability to constrain their physical properties (for detailed reviews see e.g. Beltrán & de Wit 2016; Zhao et al. 2020). However, recent advancements in interferometry, allowing for sub-arcsecond resolution at (sub-)millimeter wavelengths, have led to a turning point for high-mass star formation studies, as a growing number of disks around O- and B-type stars are detected (e.g. Patel et al. 2005; Sánchez-Monge et al. 2013; Johnston et al. 2015; Ilee et al. 2016; Cesaroni et al. 2017; Csen-geri et al. 2018; Ginsburg et al. 2018; Ilee et al. 2018a; Girart et al. 2018; Moscadelli et al. 2019; Maud et al. 2019; Johnston et al. 2020b; Añez-López et al. 2020). A disk-mediated accretion process seems to be a necessity for the formation of higher-mass objects. During this process disks themselves may become unstable due to their high densities. A disk undergoing a Keplerian rotation is prone to fragmentation when self-gravity and thermal pressure are no longer in equilibrium (Toomre 1964). Using 3D radiation-hydrodynamic simulations (for a detailed modeling description see Oliva & Kuiper 2020), Ahmadi et al. (2019) study the fragmentation of a massive disk and the "observability" of such fragmentation given a variety of source inclinations and distances. This study reveals that a Toomre-unstable disk fragments on scales below 500 AU, where each fragment hosts its own smaller scale accretion disk that is best resolved in the synthetic Atacama Large Millimeter/submillimeter Array (ALMA) observations reaching an angular resolution of ~80 mas at a distance of 800 pc (corresponding to a spatial resolution of ~60 AU). The authors also find that the fragments can also be resolved with IRAM NOrthern Extended Millimeter Array (NOEMA) with $0\farcs4$ resolution if the disk is at an inclination of 10° or 30° at 800 pc. Such disk fragmentation around high-mass stars has only recently been observed using high-resolution ALMA observations revealing spiral structures within the disk (Maud et al. 2019; Johnston et al. 2020a) and secondary components (Ilee et al. 2018b).

[1] $\tau_{KH} = \frac{GM^2}{2RL}$, where $G$ is the gravitational constant, $M, R$ and $L$ are the mass, radius and the luminosity of the protostar.

In order to address the open questions in high-mass star formation studies, namely the underlying physics and chemistry during formation and fragmentation of disks along with the infall and outflow processes, we observed 20 high-mass star forming regions (HMSFRs) at 1.37 mm within the framework of the IRAM NOEMA large program "CORE"[2] (Beuther et al. 2018). The CORE sample consists of northern HMSFRs that are otherwise mostly or entirely inaccessible to the southern hemisphere observatories (e.g. ALMA). The program provides a unique understanding of the physical and chemical properties of the observed regions both with detailed case studies (e.g. Ahmadi et al. 2018; Gieser et al. 2019; Cesaroni et al. 2019; Bosco et al. 2019; Mottram et al. 2020; Olguin et al. 2020) and statistical studies of larger samples (Beuther et al. 2018; Ahmadi in prep.; Gieser et al. 2021). Amongst our sample, AFGL 2591 is one of the most luminous HMSFRs with $L$ ~$2{\times}10^5$ $L_\odot$ at a distance of 3.33 kpc (Rygl et al. 2012). It is in the direction of the Cygnus X star-forming complex and it is associated with a large-scale (>1′~1 pc), high-velocity ($\Delta V$ ≥30 km/s) molecular outflow reported by Bally & Lada (1983). Early Very Large Array (VLA) observations show that AFGL 2591 is a complex cluster of multiple sources with independent HII regions (Campbell 1984), with subsequent observations five distinct sources in the region. The brightest infrared source, VLA 3, often referred to as AFGL 2591 itself, is a hot core that is responsible for the cone-shaped reflection nebula that is seen in Gemini $K'$ band images (for an overview of the region see Gieser et al. (2019), their Fig. 1). At $0\farcs4$ resolution and 1.37 mm, VLA 3 is observed to be a single core (Beuther et al. 2018) with a mass of ~40 $M_\odot$ (Sanna et al. 2012). As aforementioned, VLA 3 is part of the AFGL 2591 cluster with four other sources that are prominent at centimeter wavelengths, namely the HII regions VLA 1 and VLA 2 (Trinidad et al. 2003), a B9 type star named VLA 4, and VLA 5 that is associated with an infrared source (Johnston et al. 2013).

The idea that AFGL 2591-VLA 3, hereafter referred to as VLA 3 for convenience, may be fragmenting has been put forward by a number of studies. Specifically, observations of $H_2O$ masers using the Very Long Baseline Interferometry (VLBI) towards the source reveal the existence of three distinct maser clusters (Sanna et al. 2012). Additionally, multiple outflow directions were observed in a previous CORE study using CO, SiO, and SO emission. This is indicative of either a disk wind or fragmentation below 1400 AU that the 1.37 mm observations did not resolve, with each fragment powering individual outflows (Gieser et al. 2019).

In this paper, we present a new set of high spatial resolution NOEMA observations as an extension of the CORE large program carried out at 843 µm towards VLA 3. In the spectral setup, we included highly excited molecular lines in order to be able to probe the gas kinematics in the immediate vicinity of the protostar. The data reveal fragmentation at scales of ~800 AU. The structure of this study is as follows: we describe the observations and data calibration in Section 2 and present the analysis of the continuum and line emission maps in Section 3. This is followed by a discussion in Section 4 and a summary of our results in Section 5.

## 2. Observations

Observations targeting VLA 3 ($\alpha$(J2000)=$20^h29^m24^s.8$, $\delta$(J2000)=+40°11′$19\farcs6$) were carried out using the IRAM NOEMA interferometer in February 2018. Five antennas equipped with the Band 4 receiver (275-373 GHz) were used in the most extended (A) configuration in order to achieve the highest spatial resolution possible. The LO frequency was 360.995 GHz (840 µm). Using the new correlator, PolyFiX, spectra between 353.4-357.1 GHz were taken at a spectral resolution of 2 MHz, corresponding to a velocity resolution of 1.6 km/s at 356 GHz. To obtain a higher spectral resolution for a selected set of strong lines, we placed five additional spectral windows (for each polarization) with 62.5 kHz spectral (0.05 km/s velocity) resolution. The observations were conducted in dual polarization and single-sideband mode.

Two quasars, 2013+370 and 2037+511, were used for amplitude and phase calibration. The quasar 3C84 was used for bandpass calibration, while the absolute flux calibration was done using a bright star (MWC349). The uncertainty in the flux calibration is ~20%. The total on-source time was 7.9 hours with precipitable water vapor (pwv) < 2 mm. The data reduction was performed using the Continuum and Line Interferometer Calibration[3] (`CLIC`), part of the `GILDAS` software developed by the Institut de Radioastronomie Millimétrique (IRAM). We extracted the 843 µm continuum emission from the line-free channels in the two low-resolution spectral units (L01 and L02) of PolyFiX. For the lines observed with 2 MHz resolution, the continuum emission was subtracted prior to imaging. Continuum subtraction for the lines covered by the high-spectral resolution windows was done using the `uv_baseline` method of the `MAPPING` package within `GILDAS`.

We cleaned and restored each individual map using the `MAPPING` package of `GILDAS` where we used the Clark algorithm. The continuum emission was, firstly, cleaned with no clean-box. Once we had obtained a clean continuum map, we used a clean-box around the detected emission from the source and a robust parameter of 0.1 to achieve the highest possible spatial resolution. We interactively cleaned and re-set the size of the clean box after each set of iterations until the cleaned flux as a function of the identified clean components converged. Self-calibration had been shown to improve the 1.3 mm images of the full CORE sample (Beuther et al. 2018). However, because of the limited signal-to-noise ratio and only 5 antennas being available in the array for Band 4 observations, it did not enhance this dataset. Therefore, we did not self-calibrate our data. The resulting synthesized beam size is 0.″19×0.″17 with a position angle of 57.9°. The noise level in the continuum emission is 6.7 mJy/beam, which was calculated in an emission-free area.

[2] https://www2.mpia-hd.mpg.de/core/Overview.html

[3] http://www.iram.fr/IRAMFR/GILDAS

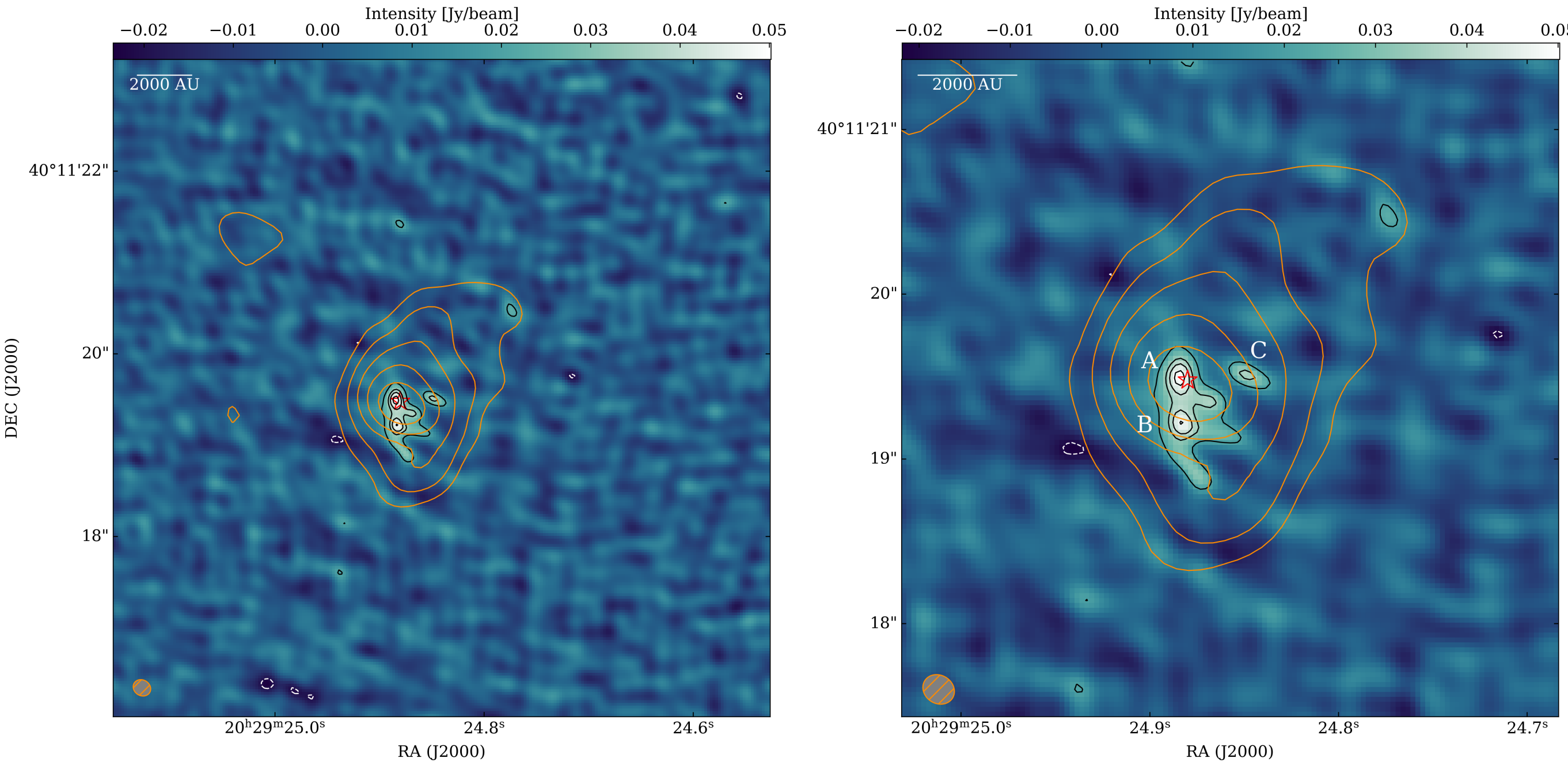

**Fig. 1.** 843 μm continuum flux density map. *Left panel*: The black contours mark 843 μm continuum emission levels of 3, 5, 6, and 7$\sigma$ with $\sigma$=6.7 mJy/beam. The white dashed contours mark the negative emission levels of 3$\sigma$. The orange contours mark the 5, 10, 20, 40, and 80$\sigma$ continuum emission levels at 1.3 mm ($\sigma$ = 0.59 mJy/beam) in both panels. *Right panel*: A close-up view of the central 13 000 AU region. Three cores, A, B, and C, detected with ≥5$\sigma$ confidence are marked with the corresponding labels. The beam size and a scale bar of the 843 μm observations is shown in the bottom left and top left corner in each panel, respectively. The red star in both panels indicates the position of the protostar, assuming that it is located at the 1.3 mm continuum peak.

We imaged the lines using the same interactive cleaning method which we used for the continuum map with the difference that the clean-box was set based on the integrated emission after each set of iterations. We chose a higher robust parameter, 1 (i.e. closer to natural weighting), for the line emission as the sensitivity for spectral lines is lower. This resulted in a beam size of 0.″21×0.″19 with a position-angle of 47.8°. The noise level, calculated in an emission-free channel, in the low- and high-resolution spectra which were imaged using a velocity resolution of 2 and 0.5 km/s, respectively, is given individually for each map in Table 1.

**Table 1.** Properties of the lines detected in the 843 μm correlator setup.

| Molecule | Transition | Frequency [GHz] | $E_u/k_B$ [K] | $\sigma^{a,b}$ [mJy/beam] |
|---|---|---|---|---|
| HCN | 4–3 | 354.505 | 42.5 | 24 |
| HCN;v$_2$=1 | 4–3, l=1e | 354.460 | 1066.9 | 85 |
| $HC_3N$ | 39–38 | 354.697 | 340.5 | |
| $SO_2$ | $12_{4,8}$–$12_{3,9}$ | 355.046 | 110.9 | 29 |
| $SO_2$ | $15_{7,9}$–$16_{6,10}$ | 356.040 | 230.4 | |
| $HC_3N$;v$_7$=1 | 39–38, l=1f | 356.072 | 662.7 | 25 |
| HCN;v$_2$=2 | 4–3, l=2f | 356.135 | 2095.2 | |
| HCN;v$_2$=2 | 4–3, l=2e | 356.163 | 2095.2 | |
| HCN;v$_2$=1 | 4–3, l=1f | 356.255 | 1067.2 | 60 |
| $HCO^+$ | 4–3 | 356.734 | 42.8 | |
| $SO_2$ | $10_{4,6}$–$10_{3,7}$ | 356.755 | 89.8 | |

**Notes.** Frequencies and upper energy levels are taken from the Cologne Database of Molecular Spectroscopy (CDMS, Müller et al. 2005). For HCN;v$_2$=1 4–3, l=1e and $SO_2$ $12_{4,8}$–$12_{3,9}$ these data are taken from the Jet Propulsion Laboratory database (Pickett et al. 1998).
[a] The rms is calculated per velocity channel. For the vibrationally excited HCN lines the channel width is 0.5 km/s, for the other lines it is 2 km/s. [b] The rms is only given for the brightest lines which are individually imaged and presented in Fig. 4. For the remaining lines detected in the low spectral resolution band but not individually imaged, the rms noise is 3.4 mJy/beam.

## 3. Analysis and results

### 3.1. Continuum emission

CORE observations towards VLA 3 at 1.37 mm provides an insight to the chemical and physical structure of its envelope using continuum and line emission maps with an angular resolution of 0.″4 which corresponds to 1400 AU at 3.33 kpc (Gieser et al. 2019). In Figure 1, we show the 843 μm continuum emission where we resolve the disk within this envelope with an angular resolution of 0.″2 arcseconds (∼700 AU). The 843 μm emission is concentrated near the center of the 1.37 mm emission and shows fragmentation.

We identify two main cores, A and B, at the center of the map detected with >5$\sigma$ confidence. The morphology of the 843 μm continuum emission is elongated with core A and B separated by ∼800 AU. We tentatively detect (∼5$\sigma$) another companion core, core C, west of core A. The separation between core C and core A is ∼1400 AU, similar to the distance between the G11.92-0.61 MM1a and MM1b (1920 AU) which are two cores in the fragmented disk of G11.92-0.61 MM 1 (Ilee et al. 2018b). Additionally, Figure 1 shows that there is another 3$\sigma$ core approximately 2″ north-west of core A, near the 5$\sigma$ contour of the 1.37 mm emission. However, we do not include this core in our analysis given the lack of confidence at the edge of the 5$\sigma$ contours of the 1.37 mm emission, and its existence remain debatable.

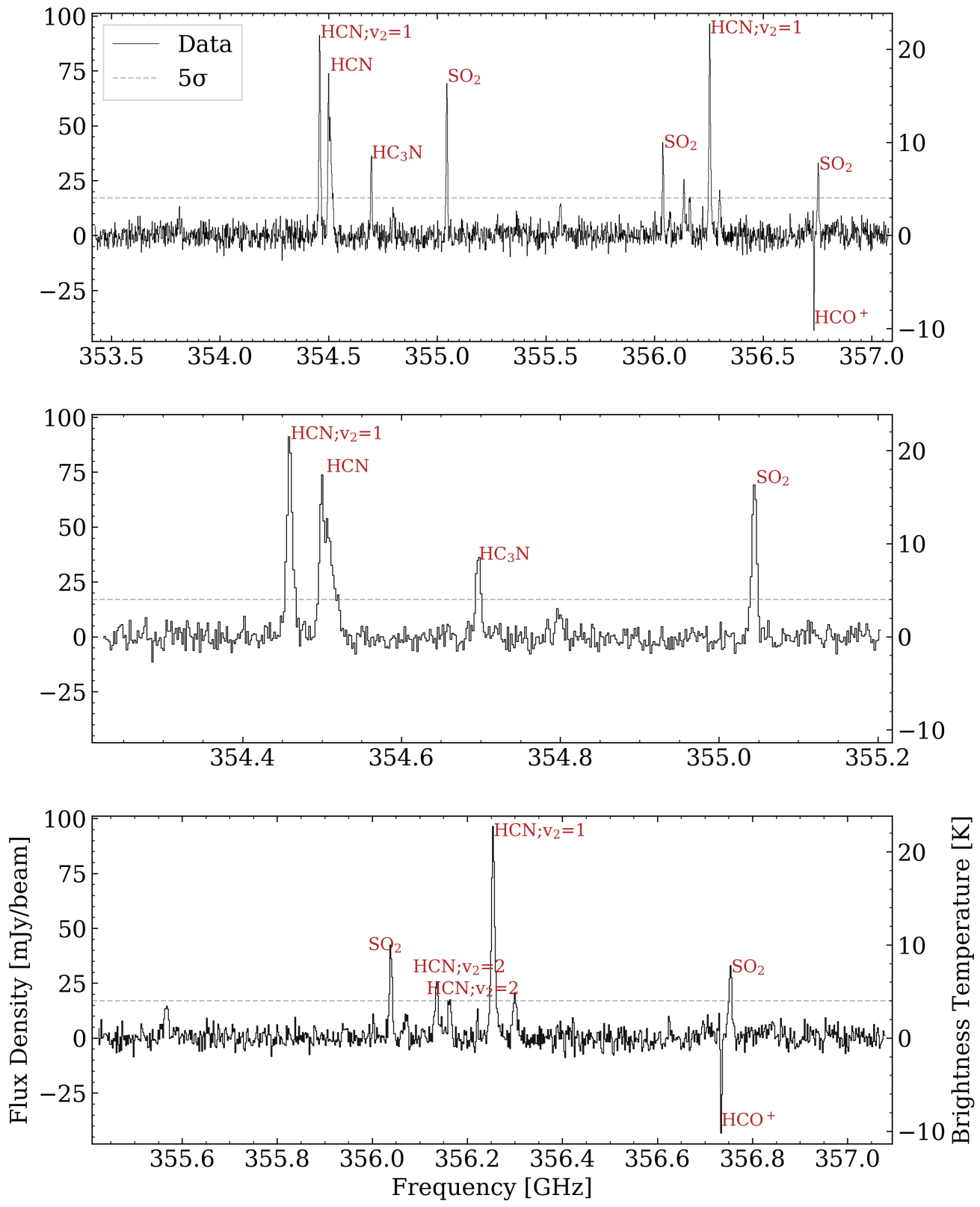

**Fig. 2.** *Upper panel:* Low spectral resolution spectrum spatially averaged over a 0″.7×0″.7 region centered at $\alpha$=20$^h$29$^m$24$^s$.876, $\delta$=+40°11′19″.216 covering cores A, B, C. *Middle panel:* Zoom-in to the frequency range from 354.2 to 355.3 GHz. *Bottom panel:* Zoom in to the frequency range from 356.4 to 357.1 GHz. The horizontal gray dashed line corresponds to the 5$\sigma$ signal-to-noise level ($\sigma$ = 3.4 mJy/beam).

For the cores identified in the 843 µm continuum emission, we calculate masses using:

$$M_{core} = \frac{S_\nu d^2 R}{B_\nu(T_d)\kappa_\nu}, \tag{1}$$

where $S_\nu$ is the flux integrated over a core, $B_\nu$ is the Planck function calculated for the dust temperature $T_d$, $d$ is the distance to the source, $R$ is to gas-to-dust mass ratio, and $\kappa_\nu$ is the dust opacity at 843 µm. From Ossenkopf & Henning (1994), we interpolate the dust opacity at 843 µm to be 1.8 cm$^2$ g$^{-1}$ (for densities of 10$^6$ cm$^{-3}$ and dust with thin ice mantles). We use a gas-to-dust mass ratio of 150 (Draine 2011). The temperature distribution of AFGL 2591 is presented in Gieser et al. (2019) (their Fig. 4). We infer the average temperatures of our cores using their $CH_3CN$ temperature map to be 200 K. Under the assumption of thermal gas-dust coupling, we use this temperature as $T_d$ to calculate core masses in Eq. 1. The sum of the masses of the three identified cores is ∼0.9 $M_\odot$. The estimate of the total mass, as well as the individual masses of the cores presented in Table 2, should be taken as a lower limit due to the interferometric filtering. The dominant errors on the mass and column density calculations stem from the uncertainties in the source distance, temperature, gas-to-dust mass ratio, and dust opacity. Accounting for these uncertainties, the calculated masses and column densities might vary within a factor of 2−4 (also reported in Beuther et al. 2018, 2021). We estimate the amount of missing flux by looking at two different studies. Firstly, in comparison to the total flux from our three cores (∼ 0.1 Jy), the total single dish flux of AFGL2591 at 850 µm observed with the James Clerk Maxwell Telescope (JCMT) is ∼8 Jy, suggesting we are filtering out more than 95% of the extended emission (Di Francesco et al. 2008). Additionally, Gieser et al. (2019) calculate the total mass within the innermost 10 000 AU of the hot core to be 6.9 $M_\odot$. Based on this, we estimate that our observations filter about 90% of the total flux. Therefore, we emphasize that our NOEMA observations are only resolving the inner disk of AFGL2591, filtering out the larger common envelope of the cores. We note also that at these high temperatures the thin dust opacities may no longer be valid due to ices evaporating off the grain surfaces. Hence, if the opacity is higher than the assumed value of 1.8 cm$^2$ g$^{-1}$, the fragment masses inferred from the 843 µm continuum data are lower than the values which are presented in Table 2.

We also calculate the column densities towards the central peaks of each core using Hildebrand (1983):

$$N(H_2) = \frac{I_\nu^{\mathrm{peak}} R}{\mu m_{\mathrm{H}} \Omega \kappa_\nu B_\nu(T_d)}, \tag{2}$$

where $I_\nu^{\mathrm{peak}}$ is the peak intensity, $\Omega$ is the beam solid angle, $\mu$ is the mean molecular weight, and $m_{\mathrm{H}}$ is the mass of the hydrogen atom. We find that the column densities towards the center of the cores are on the order of 10$^{24}$ cm$^{-2}$ which translates into a relatively substantial visual extinction of >500 mag. The column densities of the cores are also presented in Table 2.

### 3.2. Line emission and kinematics

Figure 2 shows an average spectrum extracted from a 0″.7×0″.7 area that covers the identified cores. The rotational and rovibrational HCN transitions are responsible for the brightest lines in the 843 µm spectrum. HCN becomes the most abundant N-bearing species at temperatures above ∼200 K when the water ice evaporates. This is due to the chemical reactions that involve OH, N, $N_2$, and hydrocarbons. OH forms from $H_2O$ by high-energy processes, and N is formed by destruction of $N_2$ by $He^+$. OH then rapidly reacts with N, leading to NO, which reacts with hydrocarbons and forms HCN. The key reactions of this chemistry have barriers and require temperatures of 100−200 K to activate. This high-temperature chemistry is well studied both theoretically and observationally (e.g. Rodgers & Charnley 2001; Feng et al. 2015). In addition to these, we detect multiple transitions of $SO_2$, $HC_3N$, and $HCO^+$ lines. $SO_2$ ice evaporates and becomes very abundant in the gas phase at temperatures above ∼70 K, around water snowline temperature regime (Sandford & Allamandola 1993). Amongst all, the $HCO^+$(4–3) transition is the only one detected in absorption. This is likely due to the cold $HCO^+$ gas present in the resolved-out cold outer envelope, which is dilute and, hence, more ionized than the inner disk. According to Shirley (2015), the effective density to excite a 1 K line of HCO$^+$(4–3) (for $T_{\mathrm{kin}}$ = 50 K) is $7.2 \times 10^3$ cm$^{-3}$. If the envelope around VLA 3 contains some extended volume where the density is on the order of $<10^4$ cm$^{-3}$ and the temperature is still elevated and >40 K, the extended HCO$^+$ assumption which leads to such an absorption line can become valid. However, because we filter out a large fraction of the extended emission, we can not rule out the possibility that the absorption feature could be an imaging artifact. The HCO$^+$(4–3) absorption line is blended with the adjacent $SO_2$ emission line and unresolved in the low spectral resolution spectrum. A full list of the detected molecules and their transition properties is presented in Table 1.

**Table 2.** Properties of the cores identified at 843 µm continuum.

| Core | RA (J2000) | DEC (J2000) | $I_\nu^{\mathrm{peak}}$ [mJy/beam] | $S_\nu$ [mJy] | $M_{\mathrm{core}}$ [$M_\odot$] | $N_{\mathrm{H_2}}$ [$10^{24}$ cm$^{-2}$] |
|---|---|---|---|---|---|---|
| A | $20^h29^m24^s.89$ | $+40°11'19''.43$ | 50.1 | 61.0 | 0.4 | 1.1 |
| B | $20^h29^m24^s.88$ | $+40°11'19''.24$ | 47.6 | 50.3 | 0.3 | 1.1 |
| C | $20^h29^m24^s.85$ | $+40°11'19''.46$ | 36.8 | 36.8 | 0.2 | 0.8 |

The bright ro-vibrational ($v_2$=1) HCN (4–3) lines with the excitation temperatures of ~1060 K and critical densities larger than $10^{10}$ cm$^{-3}$ probe the warm and dense material in close proximity to the protostar. We also detect higher excitation $v_2$=2 emission towards VLA 3. However, these lines have a much lower signal-to-noise ratio (~5), and hence, are difficult to image. These lines were previously detected towards AFGL2591 using the Submillimeter Array (SMA), however, due to the lower angular resolution of these data, the structure of the inner few thousand AU of the AFGL2591 disk remained unresolved (Benz et al. 2007; Veach et al. 2013).

For the further analysis we use the most prominent lines in our dataset: HCN (4–3), HCN (4–3) $v_2$=1 (both l=1e and l=1f), $HC_3N$ (39–38), and $SO_2$ ($12_{4,8}$–$12_{3,9}$). Figure 3 shows a closer look into the line emission towards the center of each core, averaged over a beam size. We note that among these transitions, the high signal-to-noise of the ro-vibrational HCN lines allowed for imaging with a higher velocity resolution of 0.5 km/s in comparison to the rest of the lines presented in Figure 3, which were imaged with a velocity resolution of 2 km/s. The l=1f transition of the vibrationally excited HCN is the strongest, reaching a peak brightness of 100 K towards core A, and 60 K towards core B. These values are much higher than the previously reported in the SMA observations of the same line of 30 K (Benz et al. 2007) and 15 K (Veach et al. 2013) observed with a lower angular resolution of 0.″6 and 3.″4 × 2.″0. The ground state HCN line is the widest and shows a dip at the source velocity towards cores A and B, indicating optically thick emission. It is also skewed and more extended towards the redshifted velocities which may be associated with the protostellar outflow. Unlike cores A and B, we do not detect vibrationally excited HCN emission towards core C. Interestingly, the ground state HCN emission and $SO_2$ emission towards core C are at equal brightness unlike cores A and B with HCN emission less self-absorbed in this line of sight due to lower column densities.

We present the moment maps (integrated intensity, intensity-weighted peak velocity, and velocity linewidth) of five selected transitions in Figure 4. All the HCN lines show the extended morphology of the source in the north-south direction, with the ground state HCN line emission being extended also in the east-west direction towards core C. The integrated emission between −10 km/s and ~1 km/s of the vibrationally excited transitions of HCN peaks between the cores A and B, although their channel maps show that these two cores can be identified separately (see Figure A.1 and A.2 ). The $SO_2$ emission is widespread, covering all three cores, and also extended north and north-east of core A. This behavior suggests that core A and its immediate vicinity is exposed to shocks. The integrated $HC_3N$ emission peaks near core A, is bright towards core B, but not detected towards core C. The spatial distribution of the $HC_3N$ emission in comparison to 1.3 mm continuum emission towards VLA 3 is discussed in detail Jiménez-Serra et al. (2012), observed using SMA with a 0.″5 resolution. Our higher resolution observations confirm the double-peak nature of the emission (see Figure A.5), even though integrated emission is brightest in between cores A and B.

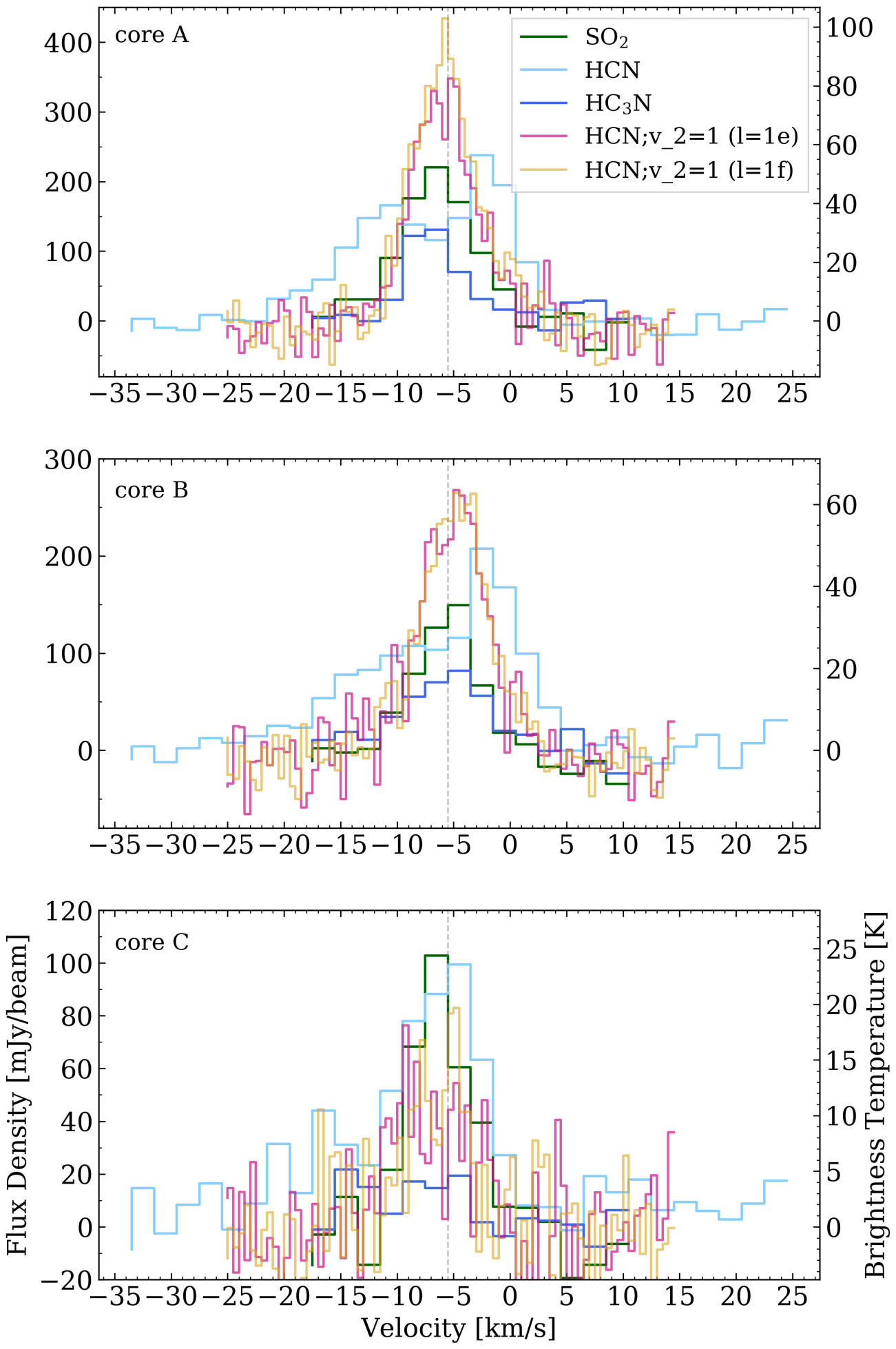

**Fig. 3.** Spectra of the five most prominent transitions towards the center of the cores A (top panel), B (middle panel), and C (bottom panel) averaged over a beam size. The vertical gray dashed line indicates the source velocity.

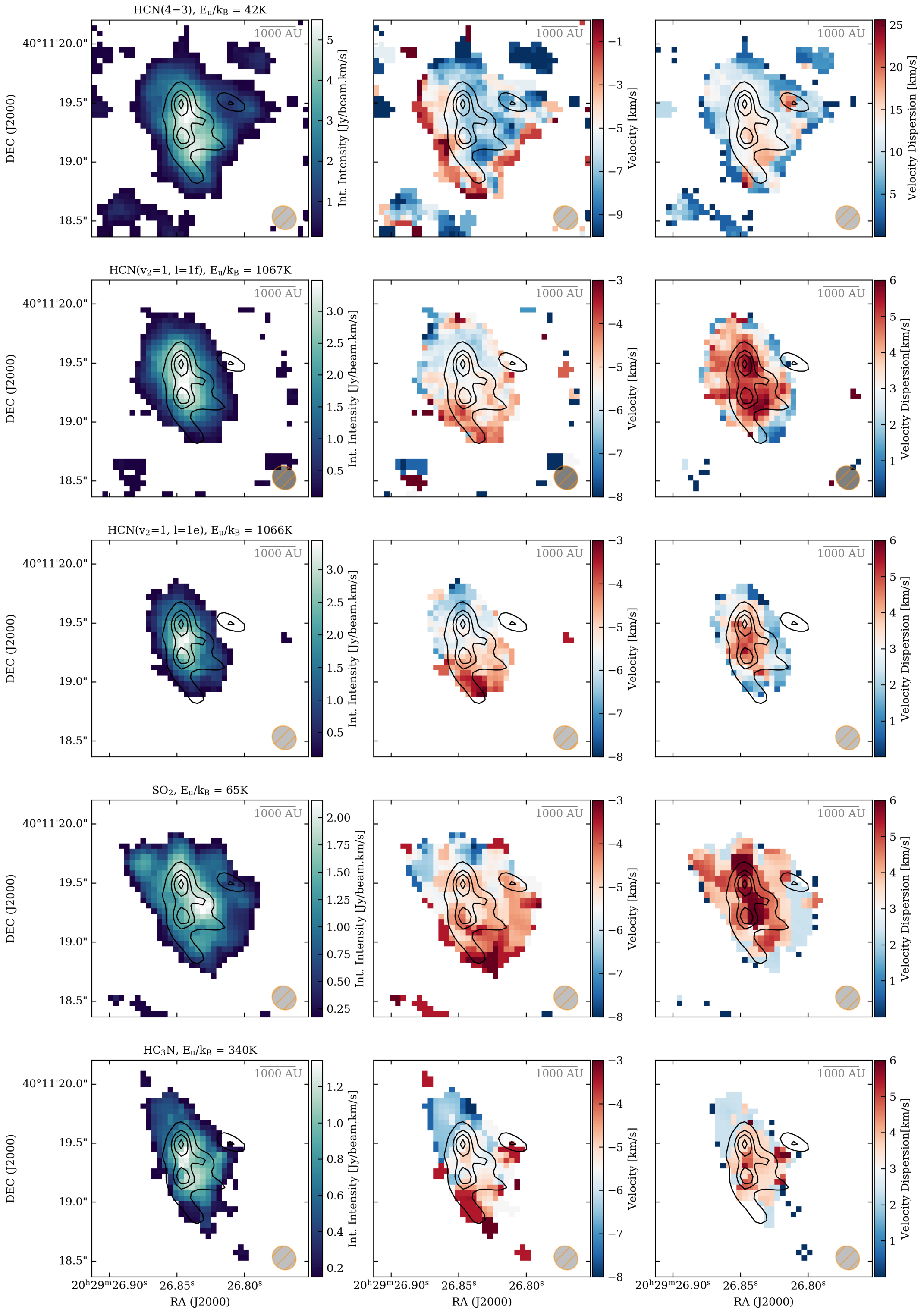

**Fig. 4.** Moment maps of the five most prominent lines in our observations. Each row represents a line and the columns, from left to right, show the integrated intensity, intensity-weighted velocity, and velocity dispersion maps. The contours in each sub-panel indicate the 3, 5, 6 and 7$\sigma$ emission levels of the 843 µm continuum emission. The beam size of the observations are shown in the bottom right corner of each sub-panel. The source velocity is −5.5 km/s.

All the peak velocity maps presented in Figure 4, except for the ground state HCN(4–3) emission, show a global northeast-southwest velocity (blue to red) gradient. The velocity structure of the HCN(4–3) emission is rather complex because it flattens near the source velocity (see Figure 3) and it is self-absorbed towards cores A and B. Therefore, the velocity structure of the disk cannot be accurately inferred from this transition. On the larger scale, the $CH_3CN$ emission observed at 1.37 mm within the framework of the lower angular resolution CORE observations shows a similar blue-red velocity gradient towards AFGL 2591 consistent with our observations (Ahmadi et al. in prep.). The similarity implies that the velocity gradient of the warm and dense gas that we are seeing at the scale of ~2000 AU may be inherited from the larger scale gas motions.

Figure 4 shows that the velocity dispersion increases towards the locations of the cores. As also can be seen from the spectra shown towards each individual core in Figure 3, the HCN(4–3) rotational ground state line exhibits large linewidths. The broad $SO_2$ linewidths extend further north of cores A and B, whereas $HC_3N$ emission shows the largest linewidths are found in between cores A and B, and between cores B and C, although core C itself is not detected in $HC_3N$ emission.

## 4. Discussion

### 4.1. Disk fragmentation and kinematics

Numerical models of disk formation (e.g. Oliva & Kuiper 2020) predict disk fragmentation and formation of companion stars and spiral-like structures. Synthetic observations created using these (Ahmadi et al. 2019) and other models (Jankovic et al. 2019) predict that the substructures of the disks are observable with interferometers such as ALMA and NOEMA. In the models discussed by Ahmadi et al. (2019), the gravitationally unstable disk fragments and forms multiple companion cores in hydrostatic equilibrium within the disk, where each core is observed to form its own disk. In our observations, we detect a single fragmented disk with three companion cores with separations of 800 and 1400 AU.

A similar observational example of a fragmenting, 1000 AU disk is AFGL 4176 (Johnston et al. 2020a). In their earlier, low resolution study (Johnston et al. 2015), the authors report an unresolved, large structure in Keplerian rotation. Their higher resolution (30 mas, Johnston et al. (2020a)) observations, however, show that the disk is fragmented into a spiral and is Toomre unstable. The eastern spiral arm also shows a condensation separated by ~700 AU from the central peak (see their Figure 2), comparable to the core separations that we see towards VLA 3.

The vibrationally excited HCN lines are good tools to investigate the hot dense gas close to the protostars and the energy levels of the two vibrationally excited HCN transitions we observed are very close (~1060 K, cf. Table 1), allowing us to average these lines and study the disk kinematics with a lower noise level (~30 mJy/beam). In Figure 5, we present the resulting intensity-weighted peak velocity map of the averaged HCN lines, $v_2$=1, l=1e and $v_2$=1, l=1f, and a position-velocity (PV) diagram obtained along the northeast-southwest velocity gradient.

In order to estimate the kinematic mass of the protostar, we fit Keplerian rotation curves to the PV diagram using the publicly available python package `KeplerFit`[4] developed by Bosco et al. (2019). `KeplerFit` uses a method developed by Seifried et al. (2016) which estimates the most extreme velocities at each radial position from the protostar, and then fits a Keplerian velocity profile to the extracted PV data points. In order to prevent fitting the unresolved emission which often results in flat velocity profiles, we refrain from fitting the velocities close to the protostar, hence, exclude the inner 700 AU (~one beam size) from the fits. By fitting the rest of the velocities down to a $5\sigma$ emission level, we infer an enclosed mass of $10.3\pm1.8\ M_\odot$. The `KeplerFit` routine assumes an inclination of 90° and because we observe only a fraction of the rotational velocity (proportional to $sin^2(i)$ where $i$ is the inclination), the mass estimates are always a lower limit (see Eq. A.3 in Bosco et al. 2019). The reduced $\chi^2$ value for the Keplerian fit is 0.47. This is lower than an expected value of 1 for a good fit, however, based on previous studies (e.g. Wang et al. 2012) we know that the AFGL 2591 disk is not in perfect Keplerian rotation.

Based on the mass-loss estimates at 1.3 cm and the IR luminosity of the region, Sanna et al. (2012) estimate the mass of VLA 3 to be in the range of 20−38 $M_\odot$. There are a few reasons for the discrepancy between these higher masses for the region and the enclosed mass we calculated assuming a Keplerian disk traced by the vibrationally excited HCN emission. Previously, using HDO and $H_2^{18}O$ observations with 0''.5 resolution, Wang et al. (2012) show that the northeast-southwest velocity gradient is a combination of sub-Keplerian rotation and a Hubble-like expansion likely to be caused by the massive outflow and stellar wind. It is also likely that there is a large amount of mass outside of the region that the vibrationally excited HCN emission traces or filtered out by the interferometric observations which would impact the rotation profile of the disk. Additionally, if there is still accretion ongoing, then the total luminosity is the sum of the intrinsic luminosity of the protostar and the accretion luminosity. Hence, the mass calculated from the dynamical fit should strictly just be compared to a mass derived from intrinsic luminosity of VLA 3.

### 4.2. Spatially coincident maser emission

$H_2O$ masers are commonly accepted as signposts of active star formation where masing occurs under conditions of high density ($>10^8$ cm$^{-3}$) and temperatures (>400 K) (Elitzur et al. 1989). These conditions can be met in protostellar disks and, also, in shocked gas due to e.g. outflows (e.g. Uscanga et al. 2010). In the case of VLA 3, Sanna et al. (2012) present a detailed study on the kinematics of $H_2O$ masers using multi-epoch Very Long Baseline Array (VLBA) observations. They find that the water maser emission is spatially distributed in three clusters within a 2000 AU region. In order to investigate the locations of these maser clusters in relation to the fragments that we detect in this study and the stability of the VLA 3 disk, in Figure 6 we present the Toomre Q map overlaid with the 843 μm continuum emission and the maser locations. The Toomre Q map is derived as a part of the sample study of the kinematics and stability of the CORE sources in 1.37 mm emission (Ahmadi et al. in prep.). For AFGL 2591, the authors use the temperatures obtained from the $CH_3CN$ emission ($T_{\rm median}$ ~130 K) and assume a 16 $M_\odot$ star in the center. The assumed mass of 16 $M_\odot$ is obtained from the Keplerian fit of the $CH_3CN$ PV diagram. This value is larger than the kinematic mass obtained from our 843 μm observations, however, as aforementioned, the 1.37 mm emission traces the extended emission that is filtered out in the 843 μm maps, hence the larger enclosed mass. As can be seen in the figure, the disk is unstable (Q < 2) to a large degree except for the immediate surroundings of core A.

The maser emission from the SE cluster coincides with the continuum peak of core A within ~0''.1. While we observe no continuum cores at the exact locations of the MW or NE clusters, there is $SO_2$ emission extending beyond core A towards the NE cluster (see Figure 4, bottom panel). Indeed, the $SO_2$ velocity linewidth in this location is broad (~6 km/s), consistent with the scenario proposed by Sanna et al. (2012), that the NE cluster is associated with a young protostar named VLA 3-N just north of our core A. The velocity of the warm dense gas emission observed towards the NE cluster, between −9 to −7 km/s, is in agreement with the velocities of the maser emission in the cluster confirming the physical connection between the warm dense gas we observe and the maser emission.

[4] `KeplerFit` is freely available on GitHub: `https://github.com/felixbosco/KeplerFit`).

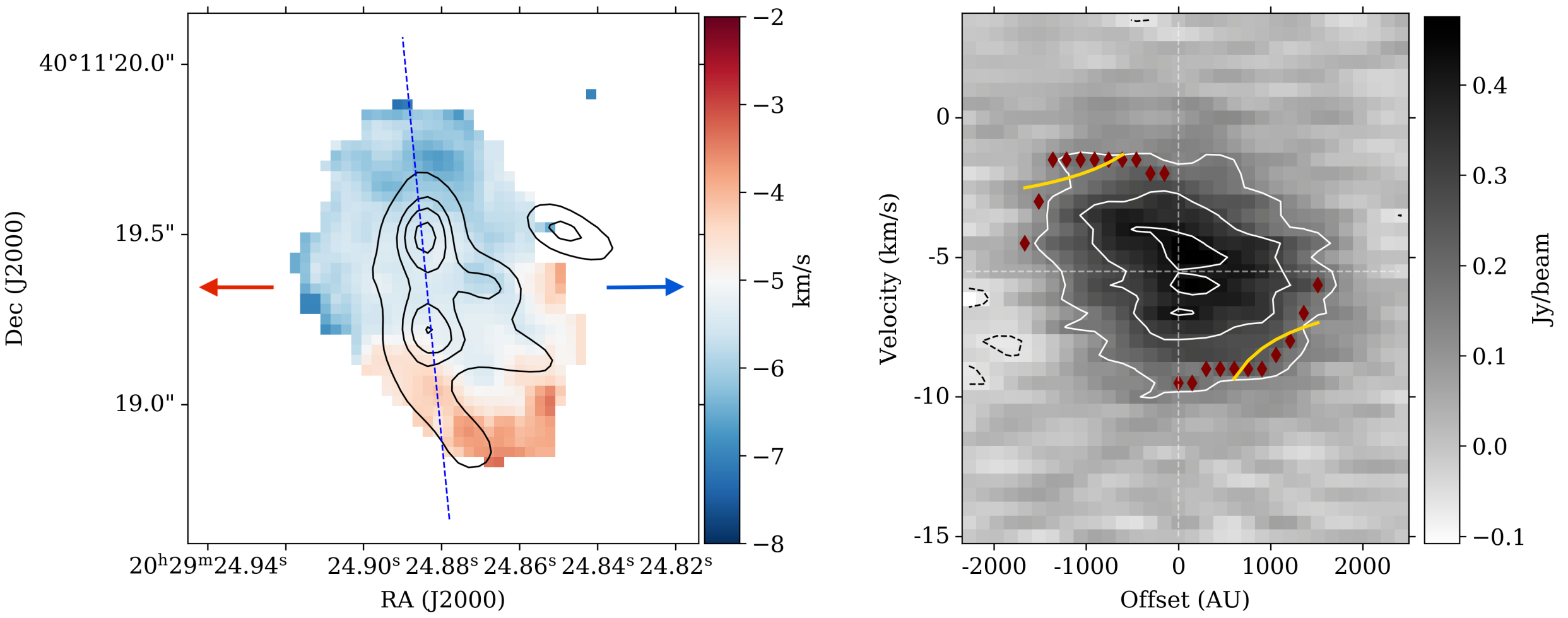

**Fig. 5.** *Left panel:* Intensity-weighted peak velocity map of the averaged vibrationally excited HCN emission. The black contours mark the 3, 5, 6, and 7$\sigma$ 843 µm continuum emission. The arrows show the direction of the large scale, east-west outflow seen in CO(2−1) emission, with blue and red colors indicating the blueshifted and redshifted line emission (see Gieser et al. 2019). *Right panel:* PV diagram for the averaged vibrationally excited HCN emission extracted from a slice along the dashed line shown in the left panel. The width of the PV slice is 9 pixels (∼0.″28). The solid, white contours mark the 5, 10, and 14$\sigma$ emission level (with 1$\sigma$ = 30 mJy/beam). The gray dashed contours mark the −2$\sigma$ negative emission levels. The red diamonds show the extreme velocities at each radius from the protostar at 5$\sigma$ emission level which are then fitted using `KeplerFit`. The fits are shown with yellow solid lines.

The location of core C, right along the axis perpendicular to the main rotation axis of the disk and between the maser clusters SE and MW, hints at another possibility for this core's origin. Along with its association with its unusually broad HCN emission (∼20 km/s, see Figure 4) and the extended $SO_2$ emission along the same direction at −4 to −6 km/s, consistent with the MW cluster velocities, suggests that core C may be a part of the east-west outflow.

An alternative scenario on the nature of the cores A and B arises when we consider the SE maser cluster and the radio continuum (Sanna et al. 2012) to be tracing the position of the protostar in VLA 3 between the cores A and B. In this case, the 843 µm continuum emission might be tracing the two edges of the disk, with the center of the disk (close to the protostar) attenuated by the outflow cavity that is seen in NIR emission (Johnston et al. 2013). However, we know that the Toomre Q values decrease towards core B, making it unlikely to have a smooth, non-fragmented disk around the protostar. The 3$\sigma$ emission around cores A and B also exhibit further sub-structure, supporting a scenario in which the disk is fragmented.

### 4.3. Disk versus core fragmentation

A number of different mechanisms, namely thermal (Jeans) or turbulent fragmentation of cores and filaments (e.g. Inutsuka & Miyama 1997; Padoan & Nordlund 2002) and disk fragmentation (e.g. Matsumoto & Hanawa 2003; Kratter & Lodato 2016; Meyer et al. 2017, 2018; Oliva & Kuiper 2020), are thought to play a role at different scales during the formation of multiple systems and clusters. Using high-resolution VLA observations, Tobin et al. (2016) characterize the multiplicity of the protostars in the low-mass star forming region Perseus molecular cloud and find that the disk fragmentation scenario is favored for separations ≲300 AU. For the scales above 1000 AU, they find that the dominating mechanism is core fragmentation (either thermal or turbulent). In the intermediate-mass regime, Palau et al. (2018) present an ALMA fragmentation study reaching a spatial resolution of 100 AU towards the OMC-1S region in Orion A. They show that the fragmentation within cores of 1000 AU of diameter can be explained by Jeans fragmentation similar to the one taking place at larger scales of 0.1 pc (similar to, e.g. Palau et al. 2015; Beuther et al. 2018).

Fragmentation in the high-mass regime below 1000 AU is poorly studied with only a handful of observational evidence for disk fragmentation (e.g. Ilee et al. 2018a; Maud et al. 2019; Johnston et al. 2020a). The initial CORE observations of VLA 3 reveal 3 cores at 1.37 mm continuum emission with mean separations of ∼15 000 AU (Beuther et al. 2018). This study shows that the fragmentation properties (mean separation and fragment masses) are in agreement with thermal Jeans fragmentation as also reported in previous studies (e.g. Gutermuth et al. 2011; Palau et al. 2015). With the current 843 µm observations, we are now looking at scales much below core fragmentation (<1000 AU) where the gravitational instabilities in protostellar disks play a vital role. Using the density ($n$ = $10^7$ $cm^{-3}$) and temperature ($T$ = 250 K) values for AFGL 2591 at ∼1000 AU reported in Palau et al. (2014), consistent with Gieser et al. (2021), we infer a Jeans mass of ∼6 $M_\odot$ and a Jeans length of ∼7000 AU. Both the core masses and separations which we calculate from the 843 µm continuum emission are an order of magnitude lower than what we would expect from the fragmentation of the parental core at 1000 AU scales, and hence, inconsistent with the Jeans fragmentation scenario. Our fragments located within the Toomre unstable disk of VLA 3, separated by ∼800 AU, suggest disk fragmentation below 1000 AU. However, we acknowledge that the Toomre Q map is produced under the assumption of Keplerian rotation, which in the case of VLA 3 was shown to be a more complex picture due to the massive outflow powered by the protostar (Wang et al. 2012).

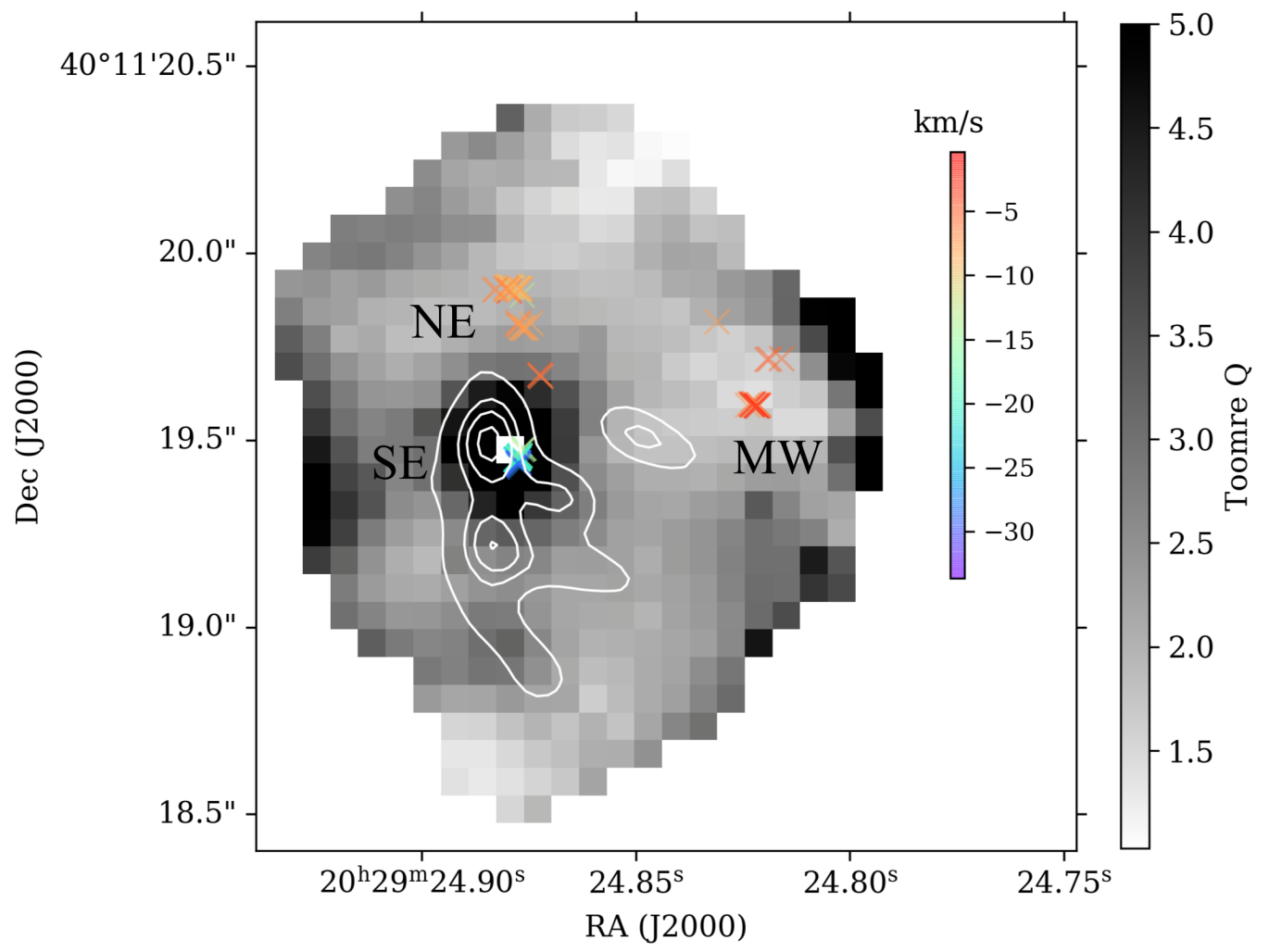

**Fig. 6.** Toomre Q map taken from Ahmadi et al. in prep. overlaid with the 843 μm continuum contours in white (3, 5, 6, and 7$\sigma$) and the water maser locations from Sanna et al. (2012). The water masers are shown with colored crosses where the color represents their radial velocity and for their locations, we use the same naming convention northeast (NE), southeast (SE), and middle-west (MW) as Sanna et al. (2012). The blanked pixel in the Toomre Q map near core A is the assumed location of the protostar.

## 5. Summary

We presented high angular resolution (0.″19 or ~700 AU) NOEMA observations towards AFGL 2591-VLA 3, conducted with the new high frequency band (Band 4) at 843 μm. The observations, for the first time, reveal fragmentation below ~1000 AU toward the AFGL 2591-VLA 3 hot core. We identified three distinct low-mass (<1 $M_{\odot}$) cores: A, B, and C. Core C is tentatively detected in the continuum emission with 5$\sigma$ confidence level, but prominent in HCN(4–3) rotational ground state and $SO_2$ line emission.

Our line emission data suggests a rotating disk-like structure with a northeast-southwest velocity gradient. Assuming Keplerian rotation for the disk, we infer a kinematic mass of 10.3±1.8 $M_{\odot}$. This velocity gradient with a similar rotation axis is also seen in previous NOEMA observations at 1.37 mm where a larger scale structure is recovered (Ahmadi et al. in prep.), suggesting that the inner disk kinematics resolved in this study at 843 μm is inherited from the larger scale gas motions.

The locations of the cores B and C are consistent with the low Toomre Q values across the disk which is largely unstable against gravitational fragmentation (Ahmadi et al. in prep.). We find that the velocities of the $H_2O$ maser clusters NE and MW are consistent with the velocities of the line emission in our observations. More specifically, we observe broad $SO_2$ and HCN emission, which is indicative of pre- or protostellar activity, towards the locations of these clusters at their radial velocities ranging between −2 km/s and −10 km/s.

Our results support a formation scenario for the high-mass objects through disk accretion where disk themselves may fragment to form companion cores. Future observations with higher sensitivity will provide an insight to whether the identified cores within the main protostellar disk have formed their own disks, as predicted by recent theoretical work.

*Acknowledgements.* We thank the anonymous referee for their time and valuable comments which improved this manuscript. This work is based on observations carried out under project number L14AB with the IRAM NOEMA Interferometer. IRAM is supported by INSU/CNRS (France), MPG (Germany) and IGN (Spain). The authors thank Alberto Sanna for sharing the locations of the water masers. SSuri and HB acknowledge support from the European Research Council under the Horizon 2020 Framework Program via the ERC Consolidator Grant CSF-648405. ASM is supported by the Collaborative Research Centre 956, subproject A6, funded by the Deutsche Forschungsgemeinschaft (DFG), project ID 184018867. RK acknowledges financial support via the Emmy Noether and Heisenberg Research Grants funded by the German Research Foundation (DFG) under grant no. KU 2849/3 and 2849/9. AP acknowledges financial support from the UNAM-PAPIIT IN111421 grant, the Sistema Nacional de Investigadores of CONACyT, and from the CONACyT project number 86372 of the 'Ciencia de Frontera 2019' program, entitled 'Citlalcóatl: A multiscale study at the new frontier of the formation and early evolution of stars and planetary systems', México. DS acknowledges support by the Deutsche Forschungsgemeinschaft through SPP 1833: "Building a Habitable Earth" (SE 1962/6-1). REP is supported by a Discovery grant from NSERC.

[1] Max Planck Institute for Astronomy, Königstuhl 17, 69117 Heidelberg, Germany
e-mail: suemeyye.suri@univie.ac.at
[2] Leiden University, Niels Bohrweg 2, 2333 CA Leiden, Netherlands
[3] I. Physikalisches Institut, Universität zu Köln, Zülpicher Str. 77, D-50937, Köln, Germany
[4] IRAM, 300 rue de la Piscine, Domaine Universitaire, 38406 St.-Martin-d'Hères, France
[5] INAF, Osservatorio Astrofisico di Arcetri, Largo E. Fermi 5, I-50125 Firenze, Italy
[6] Laboratoire d'astrophysique de Bordeaux, Univ. Bordeaux, CNRS, B18N, allèe Geoffroy Saint-Hilaire, 33615 Pessac, France
[7] Academia Sinica Institute of Astronomy and Astrophysics, No.1, Sec. 4, Roosevelt Rd, Taipei 10617, Taiwan, Republic of China
[8] CAS Key Laboratory of FAST, National Astronomical Observatories, Chinese Academy of Sciences, Beijing 100101, People's Republic of China
[9] National Astronomical Observatory of Japan, National Institutes of Natural Sciences, 2-21-1 Osawa, Mitaka, Tokyo 181-8588, Japan
[10] School of Physics & Astronomy, University of Leeds, Leeds LS2 9JT, UK
[11] UK Astronomy Technology Centre, Royal Observatory Edinburgh, Blackford Hill, Edinburgh EH9 3HJ, UK
[12] Zentrum für Astronomie der Universität Heidelberg, Institut für Theoretische Astrophysik, Albert-Ueberle-Straße 2, 69120 Heidelberg, Germany
[13] Institut für Astronomie und Astrophysik, Universität Tübingen, Auf der Morgenstelle 10, D-72076 Tübingen, Germany
[14] INAF, Osservatorio Astronomico di Cagliari, Via della Scienza 5, I-09047, Selargius (CA), Italy
[15] Astrophysics Research Institute, Liverpool John Moores University, Liverpool, L3 5RF, UK
[16] European Southern Observatory, Karl-Schwarzschild-Str. 2, D-85748 Garching, Germany
[17] Instituto de Radioastronomía y Astrofísica (IRyA), UNAM, Apdo. Postal 72-3 (Xangari), Morelia, Michoacán 58089, Mexico
[18] Max-Planck-Institut für Astrophysik, Karl-Schwarzschild-Str. 1, 85748 Garching, Germany
[19] Department of Physics and Astronomy, McMaster University, Hamilton, Ontario L8S4M1, Canada
[20] School of Physics & Astronomy, Cardiff University, Queen's building, The parade, Cardiff, CF24 3AA, UK
[21] Department of Chemistry, Ludwig Maximilian University of Munich, Butenandtstr. 5-13, House F, D-81377 Munich, Germany
[22] Centre for Astrophysics and Planetary Science, University of Kent, Canterbury, CT2 ,7NH, UK
[23] Max-Planck-Institut für Radioastronomie (MPIfR), Auf dem Hügel 69, 53121 Bonn, Germany
[24] Universidad Autonoma de Chile, Avda Pedro de Valdivia 425, Providencia, Santiago de Chile, Chile

## Appendix A: Channel maps

In this section, we present the channel maps of the most prominent molecules in our study.

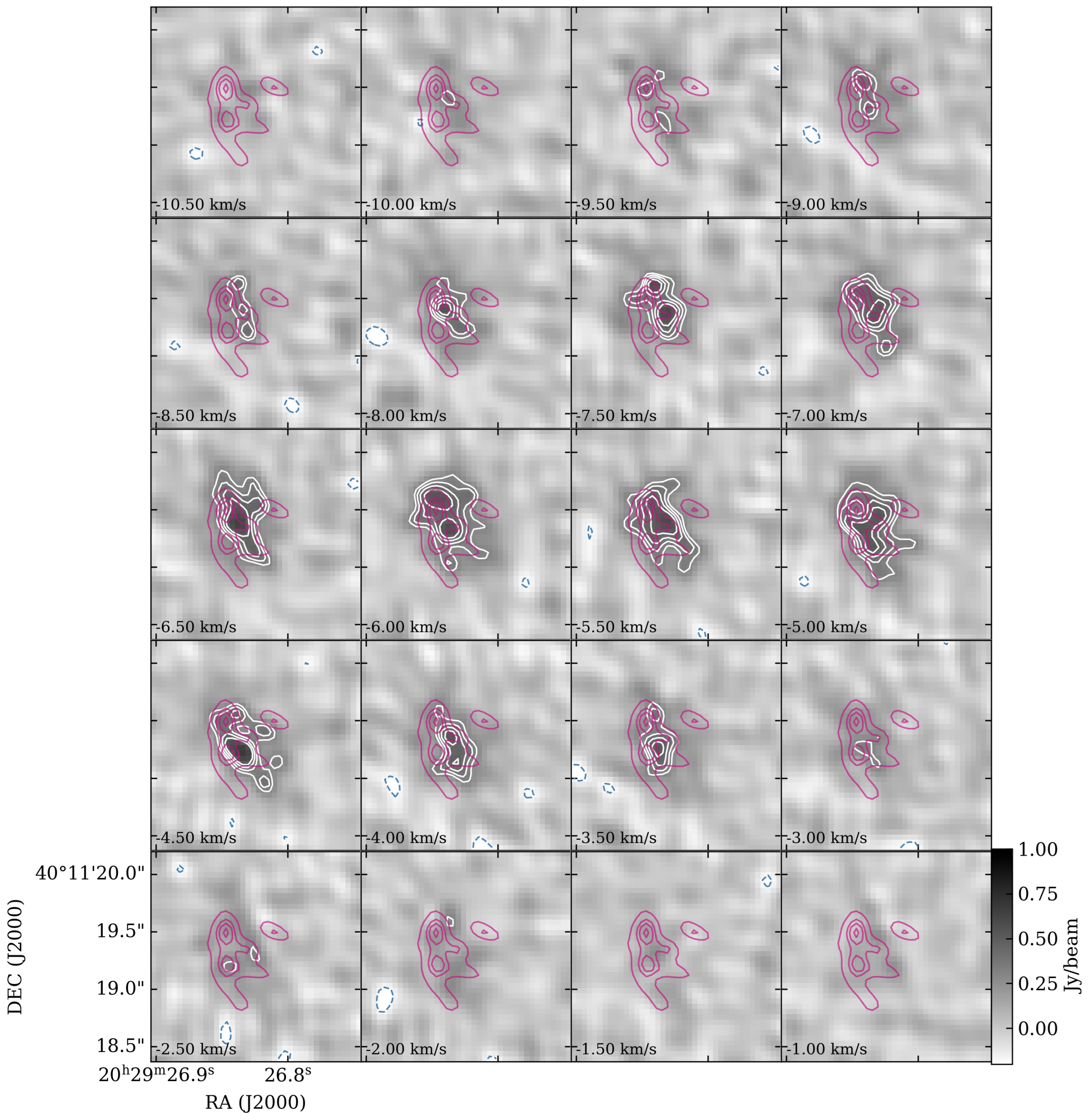

**Fig. A.1.** Channel maps of HCN $v_2$=1 emission with the emission levels of 5, 6, 7, and 8$\sigma$ are marked with white contours. The magenta contours mark the 3, 5, 6, and 7$\sigma$ 843 μm continuum emission. The blue dashed contours indicate the $-3\sigma$ line emission.

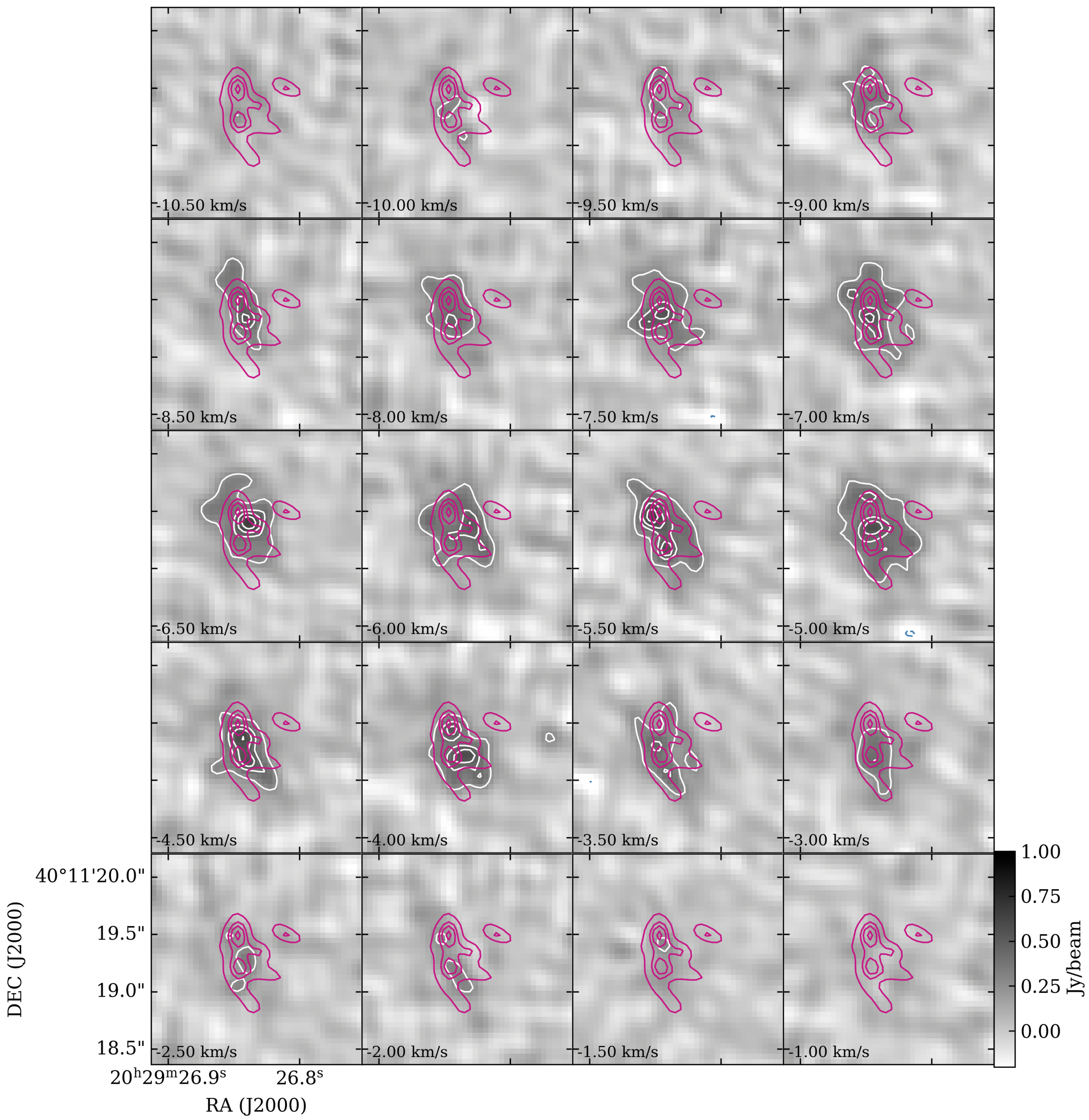

**Fig. A.2.** Channel maps of HCN l=1e emission with the emission levels of 3, 5, 6, 7, and 8$\sigma$ are marked with white contours. The magenta contours mark the 3, 5, 6, and 7$\sigma$ 843 μm continuum emission. The blue dashed contours indicate the $-3\sigma$ line emission.

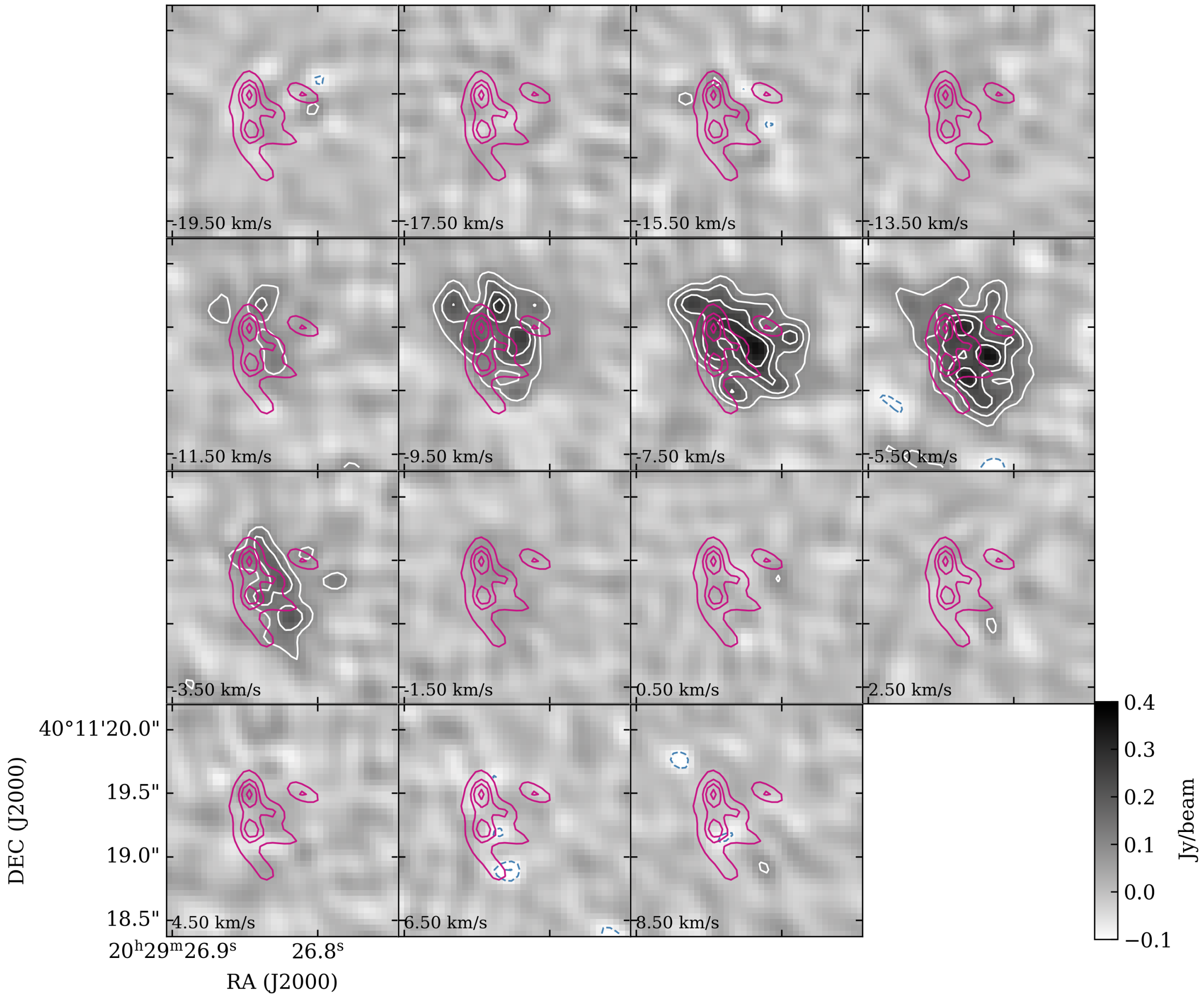

**Fig. A.3.** Channel maps of $SO_2$ emission with the emission levels of 3, 5, 6, and 7$\sigma$ are marked with white contours. The magenta contours mark the 3, 5, 6, and 7$\sigma$ 843 µm continuum emission. The blue dashed contours indicate the $-3\sigma$ line emission.

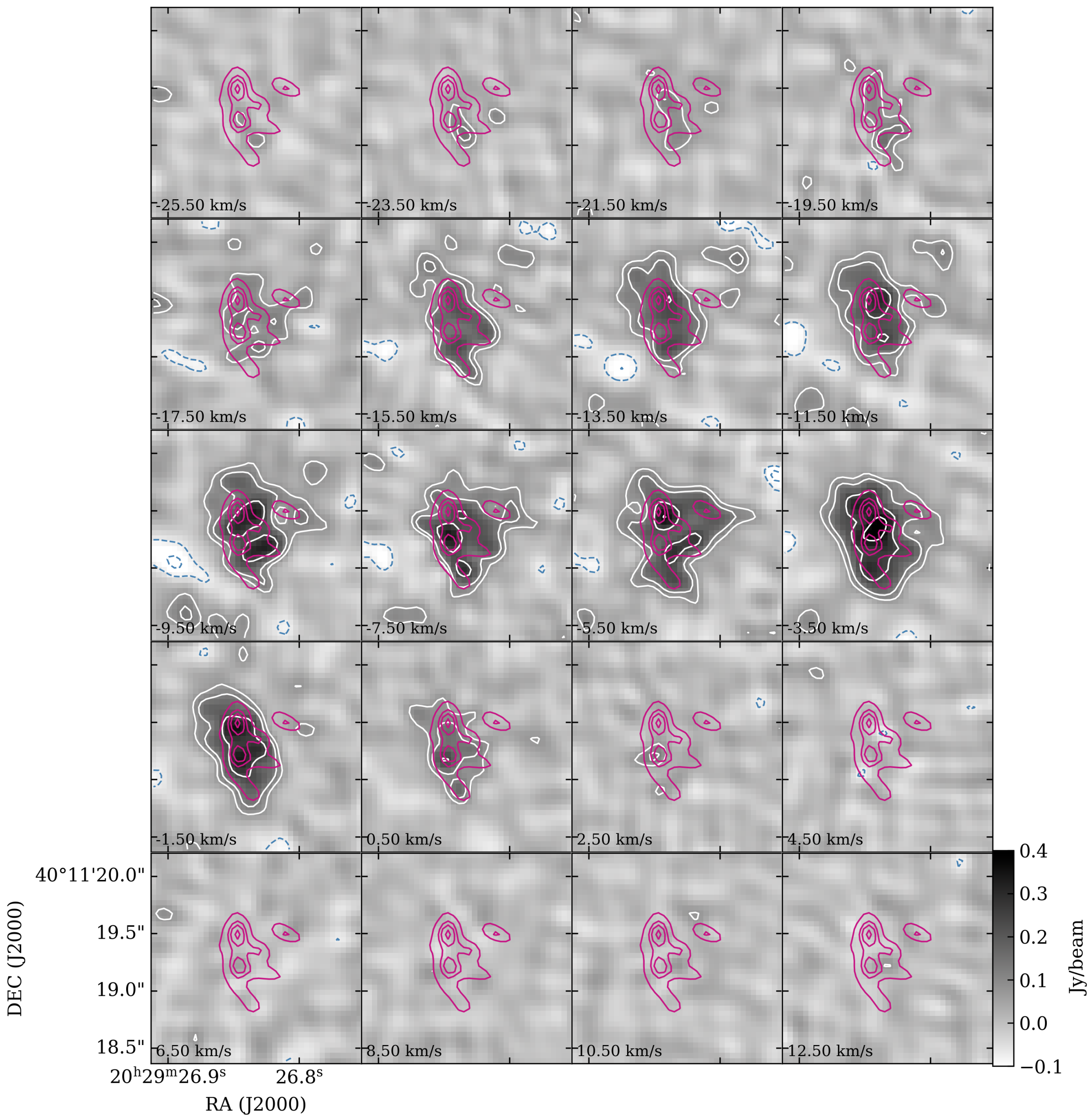

**Fig. A.4.** Channel maps of HCN 4–3 emission with the emission levels of 3, 5, 10, and 15$\sigma$ are marked with white contours. The magenta contours mark the 3, 5, 6, and 7$\sigma$ 843 µm continuum emission. The blue dashed contours indicate the $-3\sigma$ line emission.

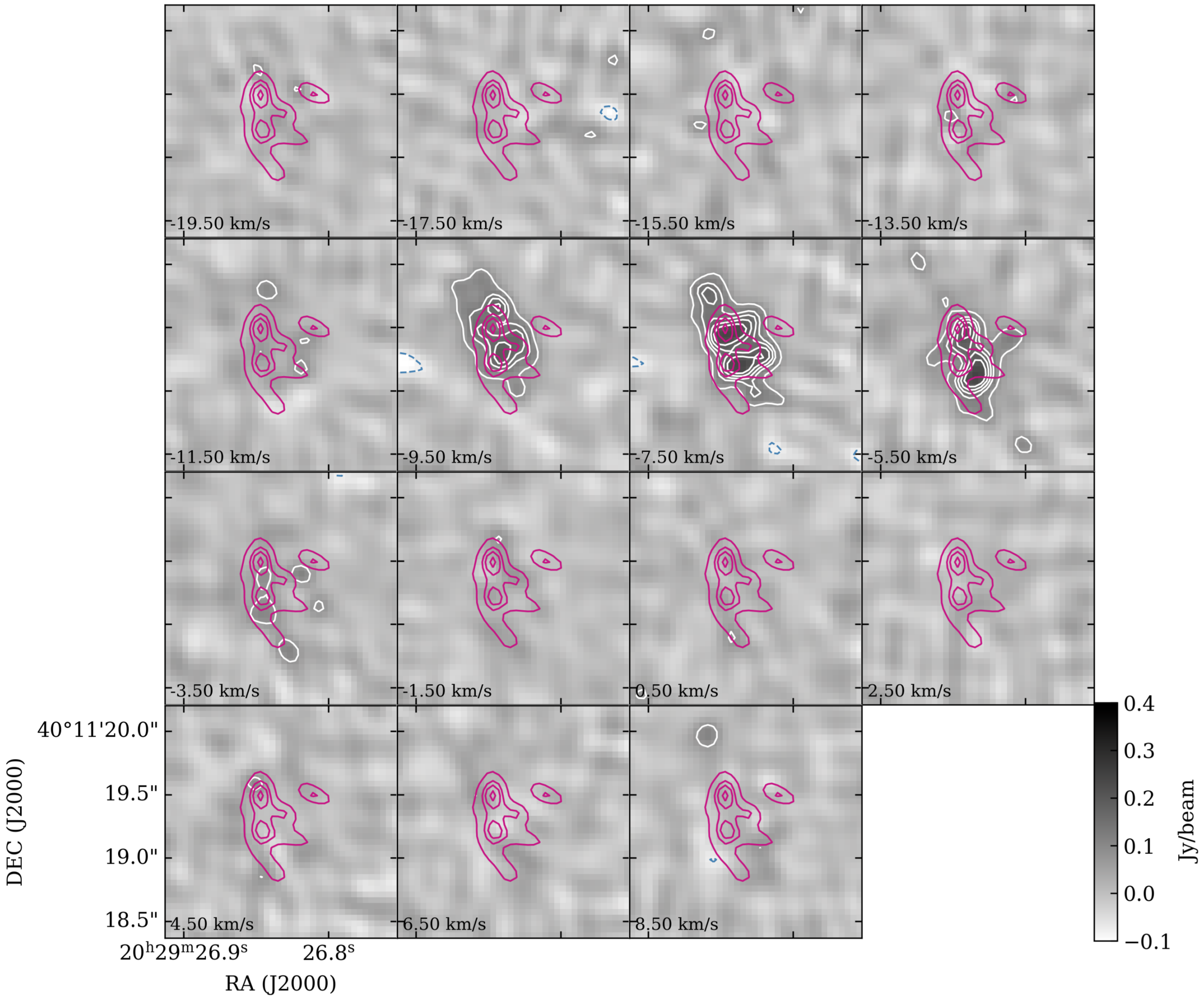

**Fig. A.5.** Channel maps of $HC_3N$ emission with the emission levels of 3, 5, 6, and 7$\sigma$ are marked with white contours. The magenta contours mark the 3, 5, 6, and 7$\sigma$ 843 µm continuum emission. The blue dashed contours indicate the $-3\sigma$ line emission.